\documentclass[preprint,12pt]{elsarticle}
\usepackage{lineno,hyperref}
\usepackage{graphicx}% Include figure files
\usepackage{dcolumn}% Align table columns on decimal point
\usepackage{bm}% bold math
\usepackage{cancel,amssymb,amsmath}
\usepackage{enumitem}
\usepackage[T1]{fontenc}
\usepackage{lmodern}
\usepackage[none]{hyphenat}

\journal{Chemical Physics Letters}

\begin{document}
\sloppy
\begin{frontmatter}

\title{Macroscopic production of highly nuclear-spin-polarized molecules from IR-excitation and photodissociation of molecular beams}


\author[a,b]{C. S. Kannis}
%\ead[url]{c.kannis@fz-juelich.de}

\author[b,c]{T. P. Rakitzis\corref{correspondingauthor}}
\cortext[correspondingauthor]{Corresponding author}
\ead[url]{ptr@iesl.forth.gr}

\address[a]{Institute for Nuclear Physics, Forschungszentrum J\"ulich, 52425 J\"ulich, Germany}
\address[b]{University of Crete, Department of Physics, Herakleio, Greece}
\address[c]{Foundation for Research and Technology Hellas, Institute of Electronic Structure and Laser, N. Plastira 100, Heraklion, Crete, Greece, GR-71110}

\begin{abstract}
Pure, highly nuclear-spin-polarized molecules have only been produced with molecular beam-separation methods, with production rates up to ${\sim}3{\times}10^{12}$ s\textsuperscript{-1}. Here, we propose the production of spin-polarized molecular photofragments from the IR-excitation and photodissociation of molecular beams, with production rates approaching the tabletop-IR-laser photon fluxes of $10^{21}$ s\textsuperscript{-1}. We give details on the production of spin-polarized molecular hydrogen and water isotopes, from formaldehyde and formic acid beams, respectively. Macroscopic quantities of these molecules are important for NMR signal enhancement, and for the needs of a nuclear fusion reactor, to increase the D-T or D-\textsuperscript{3}He unpolarized nuclear fusion cross section by ${\sim}50{\%}$.
\end{abstract}

\begin{keyword}
nuclear-spin-polarized molecules\sep molecular beams\sep IR excitation\sep photodissociation\sep molecular hyperpolarization
\MSC[2010] 00-01\sep  99-00
\end{keyword}

\end{frontmatter}

\section{Introduction}

The production of nuclear spin polarization is important in several fields, including the study of spin-dependent effects in particle and nuclear physics~\cite{steffens}, and in applications of solid-state hyperpolarized nuclear magnetic resonance (NMR) ~\cite{halse, merritt, weber}. However, current methods for producing highly spin-polarized molecules, such as molecular-beam separation techniques, polarization through ``brute force'' cryogenic cooling, dynamic nuclear polarization (DNP), and pulsed-laser excitation of molecular beams, have significant production and polarization limitations for many applications.

``Brute force'' cryogenic cooling of solid D\textsubscript{2} to 35 mK in a 14.5 T field, produces a polarization of only about 13\%~\cite{terhaar}. For 100\% polarized nuclei, The D-T and D-\textsuperscript{3}He nuclear-fusion reaction cross sections are enhanced by 50\% ~\cite{hupin}, and may increase the reactor efficiency by ${\sim}75{\%}$~\cite{temporal}. A GW nuclear fusion reactor will need ${\sim}10^{21}$ s\textsuperscript{-1} of 100\% polarized D to benefit fully from this effect~\cite{grigoryev,kulsrud,moir}. Therefore, the ``brute force'' method cannot produce sufficiently polarized D for this application. 

In DNP, the unpaired electrons in free radicals are highly polarized at low temperature in a magnetic field, and this polarization is transferred to the nuclei of the target sample~\cite{ardenkjaer}. This works successfully for a wide range of molecules, however the removal of the radicals, necessary for \textit{in vivo} NMR applications, is difficult and can significantly lower the sample polarization~\cite{pinon}.

Molecular-beam separation techniques, using electric or magnetic field gradients, is the only method that can produce pure, ${\sim}100{\%}$ nuclear-spin-polarized molecules, however only microscopic production rates have been demonstrated of ${\sim}3{\times}10^{12}$ for ortho-H\textsubscript{2} and ${\sim}5{\times}10^{11}$ s\textsuperscript{-1} for ortho-D\textsubscript{2}~\cite{yu}, and below ${\sim}10^{11}$ for ortho-H\textsubscript{2}O~\cite{kravchuk, vermette}, many orders of magnitude lower than required for many applications; for example, NMR applications and tests of polarized fusion require production rates of at least $10^{17} \, {\rm s^{-1}}$. 

The single-photon rovibrational excitation of molecular beams with an IR  laser pulse~\cite{rakitzis,rubio}, followed by transfer of the rotational polarization to the nuclear spin via the hyperfine interaction, has been shown to polarize highly the nuclear spins of isolated molecules. This method of hyperpolarization has been demonstrated for Cl nuclei in HCl molecules~\cite{sofikitis,sofikitis2}. However, a path for high production rates for important nuclei, such as H/D/T, was not shown; also, high polarization was not demonstrated for H isotopes via IR excitation, as the rotational polarization preferentially polarizes the more strongly coupled Cl nucleus first. The biggest problem is that only isolated spin-polarized atoms have been produced with this method, and that the production of spin-polarized molecules (such as H\textsubscript{2} isotopes) have only been performed by recombining the polarized atoms at a surface~\cite{engels}; this is a very complicated extra step, which has not been demonstrated at high production rates. 

Here, we propose improvements to the rovibrational excitation method that will allow macroscopic production rates of pure, highly-polarized molecules (${\sim}100{\%}$), at rates that can approach the photon fluxes of the IR excitation step. The improvements are motivated by the recent availability of tabletop tunable, narrow-bandwidth, high-power IR lasers, producing ${\sim}10^{21}$ ${\rm photons \, s^{-1}}$~\cite{mirov}, and are: (i) optical excitation steps that allow \textit{only} molecules with 100\% nuclear polarization to be photodissociated, providing a pure polarization spin filter, and (ii) excitation and polarization of larger molecules which can be photodissociated to yield exclusively polarized molecular photofragments, without needing to recombine polarized atoms; this is particularly important for molecules that are not IR active (such as H\textsubscript{2} isotopes), and cannot be excited directly. Specifically, we propose the IR-excitation of the special cases of formaldehyde (CH\textsubscript{2}O) or formic acid (CH\textsubscript{2}O\textsubscript{2}), which can be photodissociated to yield exclusively the  H\textsubscript{2} + CO~\cite{suits} or H\textsubscript{2}O + CO~\cite{ma} channels, respectively. The target molecules (here H\textsubscript{2} or H\textsubscript{2}O isotopes) can then be selectively trapped at a surface. Thus, we propose the production of macroscopic quantities of ${\sim}100{\%}$ nuclear-spin-polarized isotopes of hydrogen molecules (H\textsubscript{2}, HD, DT) and water molecules (H\textsubscript{2}O, D\textsubscript{2}O), which are important for the applications of polarized nuclear fusion and NMR signal enhancement. This production method will surpass conventional beam separation methods by several orders of magnitude, and may approach the photon flux of the IR excitation lasers (details given below).

Below, and in Figs.~\ref{fig:setup} and~\ref{fig:sketches}, we give details on the production of ${\sim}100\%$ nuclear-spin-polarized molecules from IR-excitation and photodissociation of CH\textsubscript{2}O and CH\textsubscript{2}O\textsubscript{2}. The steps are described generally, but specific numbers are given for the case of CH\textsubscript{2}O (for production rates of up to $10^{20}$ s\textsuperscript{-1} spin-polarized H\textsubscript{2}):

\begin{figure}
	\centering
	\includegraphics*[width=1\textwidth]{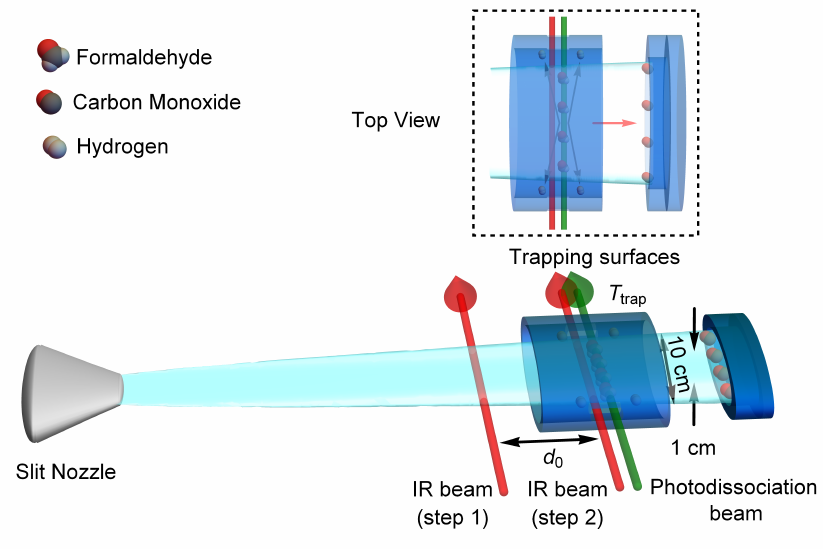}
	\caption{Experimental setup: supersonic expansion of CH\textsubscript{2}O or CH\textsubscript{2}O\textsubscript{2} gas, followed by IR excitation (step 1), time evolution for time $t_{max}=d_0/v$, then photodissociation, and trapping of polarized hydrogen or water photofragments at the cold surface. \label{fig:setup}}
\end{figure}

\begin{enumerate}[noitemsep,topsep=0pt,parsep=0pt,partopsep=0pt,leftmargin=0pt,itemindent=2em]
	\item \textit{Slit nozzle supersonic expansion.} Produces a ${\rm 10\,cm\times 1\, cm}$ jet at the laser interaction region (Fig.~\ref{fig:setup} and~\cite{nesbitt}), with CH\textsubscript{2}O density of ${\sim}2{\times} 10^{15}$ cm\textsuperscript{-3} (seeded in He) with beam velocity of ${\rm {\sim}2{\times} 10^5 \, cm \, s^{-1}}$, translational temperature of ${\sim}1$ K and rotational temperature of ${\sim}3$ K, thus cooling about half of the ortho-CH\textsubscript{2}O to the ground $J=0$ state.
	\item \textit{IR excitation (step 1).} Transition from the ground state $|\nu=0 ,\, J=0,\, m_J =0 \rangle$ to $|\nu^\prime ,\, J,\, m_J =+J \rangle$ with $\sigma_{+}$ light, and $J=1$ or 2 (Fig.~\ref{fig:sketches}). $J=1$ is used for photoproducts with total nuclear spin $I=1$ (such as ortho H\textsubscript{2} or H\textsubscript{2}O), and $J=2$ for nuclei with total nuclear spin $I=3/2$ (such as HD or DT) or $2$ (such as ortho D\textsubscript{2} or D\textsubscript{2}O), via successive two-photon absorption using STIRAP~\cite{bergmann,vitanov,bergmann2}, for which population-transfer efficiency is typically $>90\%$. The absorption cross sections are ${\sim}10^{-16}$ cm\textsuperscript{2} (for $\nu^{\prime} = \nu_2 \nu_5$ at 4581.69 ${\rm cm^{-1}}$) and the column density of the ground state is $3 {\times} 10^{16}$ cm\textsuperscript{-2} so that greater than 99\% of the $10^{21}$ s\textsuperscript{-1} IR photons are absorbed (as the 1 MHz laser linewidth is much narrower than the transition linewidth); for lower absorption cross sections, buildup cavities can be used to maximize IR absorption. 
	\item \textit{Hyperfine polarization beating.} Subsequently, we leave the system to evolve freely, transferring rotational polarization to nuclear polarization, until the population of $|\nu^\prime ,\, J,\, m_J =-J\rangle$ (for $I= 1$ or $2$) or $|\nu^\prime ,\, J, \, m_J=-J+1\rangle$ (for $I=3/2$) is maximized ($t=t_{max}$). This state is 100\% nuclear-spin polarized, due to the conservation of the total angular momentum projection along the quantization axis (hyperfine beatings under similar conditions have been demonstrated for HCl, HD, D\textsubscript{2}~\cite{sofikitis,bartlett,bartlettd2}). For CH\textsubscript{2}O, 16\% of the population is 100\% nuclear-spin-polarized and is transferred to the $m_J = - 1$ state (see Section~\ref{sec2}). The total rotational energy transfer and depolarization cross section for the $|J=1,\,K=0,\, M=1\rangle$ state in CH\textsubscript{2}O is ${\sim}3 \times 10^{-15}$ cm\textsuperscript{2}~\cite{phillips} (assuming a collision-energy-independent cross section). For collisional speeds of ${\sim}3\times 10^3 \, {\rm cm \, s^{-1}}$, the de-excitation rate is ${\sim} 10^{-11} \, {\rm cm^3 \, s^{-1}}$, which, for $10^{15}$ cm\textsuperscript{3} densities, gives a de-excitation time of ${\sim}100$ ${\rm\mu s}$; therefore, $t_{max}$ should be significantly less than 100 ${\rm\mu s}$. In addition, the velocity spread in the supersonic expansion is much less than 10\%~\cite{hutzler}, so that the blurring in the peak of a half an oscillation is less than 1\%. 
	\item \textit{Hyperfine beating stopped.} At $t=t_{max}$, the beam enters a magnetic field of ${\sim}1$ mT, which stops the hyperfine beating. A sharp magnetic field gradient is necessary to decouple nuclear spin and rotation and minimize the polarization losses. This has been shown in~\cite{kannis} for the case of H\textsubscript{2} molecules; a similar calculation for CH\textsubscript{2}O is expected to show similar behavior, and can be performed in future work.
	\item \textit{IR excitation (step 2) and photodissociation.} The transition $|\nu^\prime ,\, J,\, m_J =-J\rangle \longrightarrow |\nu^{\prime\prime} ,\, J=0,\, m_J =0\rangle$ (for $I= 1$ or $2$) or the 
	$|\nu^\prime , \, J, \, m_J = -1\rangle \rightarrow\rightarrow |\nu^{\prime\prime} , \, J=1, \, m_J = +1\rangle$ (for $I=3/2$), with $\sigma_{+}$ light, so that only the 100\% nuclear-spin-polarized state is excited to the highest level. This fully-polarized state is then exclusively photodissociated by the photolysis laser (the lower states not having enough energy to photodissociate). For CH\textsubscript{2}O column density of $3 {\times} 10^{16}$ cm\textsuperscript{-2} and a photodissociation cross section of $2.5 {\times} 10^{16}$ cm\textsuperscript{2} (Fig.~\ref{fig:CH2Odiss}), more than 99\% of the molecules are photodissociated, for laser fluxes of $2 {\times} 10^{20}$ photons cm\textsuperscript{-2} s\textsuperscript{-1}. The photodissociation cross section for higher vibrational states is shifted to the green from the UV, due to the IR absorption energy of ${\sim} 1$ eV. The photodissociation cross section of formic acid at 248 nm, for the ${\rm CO \, +\, H_2 O}$ channel, is about 1000 times smaller~\cite{singleton,maeda}, therefore a buildup cavity will be needed for the photodissociation laser, at 355 nm (for 2 IR photons) or 450 nm (for 4 IR photons), to absorb a large percentage of the photolysis light. The produced H\textsubscript{2} molecules are in $J$ states peaking at ${\sim}3$~\cite{butenhoff}, and hence a static magnetic field of ${\sim}1$ T is applied in the photodissociation region to prevent significant polarization exchange with $J$~\cite{engels18}. Notice that the H\textsubscript{2} photodissociation recoil speed is much faster than the beam speed, and therefore the H\textsubscript{2} recoils radially from the photodissociation region and is trapped at the first trapping surface (see Fig.~\ref{fig:setup}); in the contrast, the heavier CO molecules recoil slower than the beam speed, and therefore the CO follows the molecular beam and is trapped with the rest of the unphotodissociated molecules (Fig.~\ref{fig:setup}). In addition, the spatial anisotropy of the CH\textsubscript{2}O recoil can be used to direct the H\textsubscript{2} photofragments preferentially towards the trapping surfaces (and less towards the molecules beam direction)~\cite{carleton}. For CH\textsubscript{2}O\textsubscript{2} photodissociation, H\textsubscript{2}O and CO can be differentially trapped by the trapping surface temperature (as CO melts at 68 K, compared to 273 K for H\textsubscript{2}O).
	\item \textit{Cold trapping.} More than 90\% of the hydrogen or water isotopes reach trapping surfaces (that cover more than 90\% of $4\pi$ steradians), and are selectively trapped at these cold trapping surfaces with unity sticking probability~\cite{govers} (the CO mainly follows the beam direction and is trapped elsewhere, with the unphotodissociated molecules (Fig.~\ref{fig:setup})). The molecules will be trapped as a solid. Spin-polarized HD has been trapped at 4 K for many days~\cite{honig}, whereas spin-polarized H\textsubscript{2}O, produced via DNP, was shown to have lifetimes of 10 s of seconds~\cite{weber} (but pure spin-polarized H\textsubscript{2}O should have a longer lifetime).
\end{enumerate}

The product of the efficiencies of the above steps 1-6 is ${\sim}10$\%. Therefore, $10^{21}$ IR photons s\textsuperscript{-1} will produce and trap an upper limit of $10^{20}$ s\textsuperscript{-1} spin-polarized molecules (this can be scaled up with larger laser powers and slit lengths). Future experiments will determine to what extent such production rates can be approached (e.g. the depolarization rates may be higher than assumed here, which will lower the production rate).

\begin{figure}
	\centering
	\includegraphics*[width=1\textwidth]{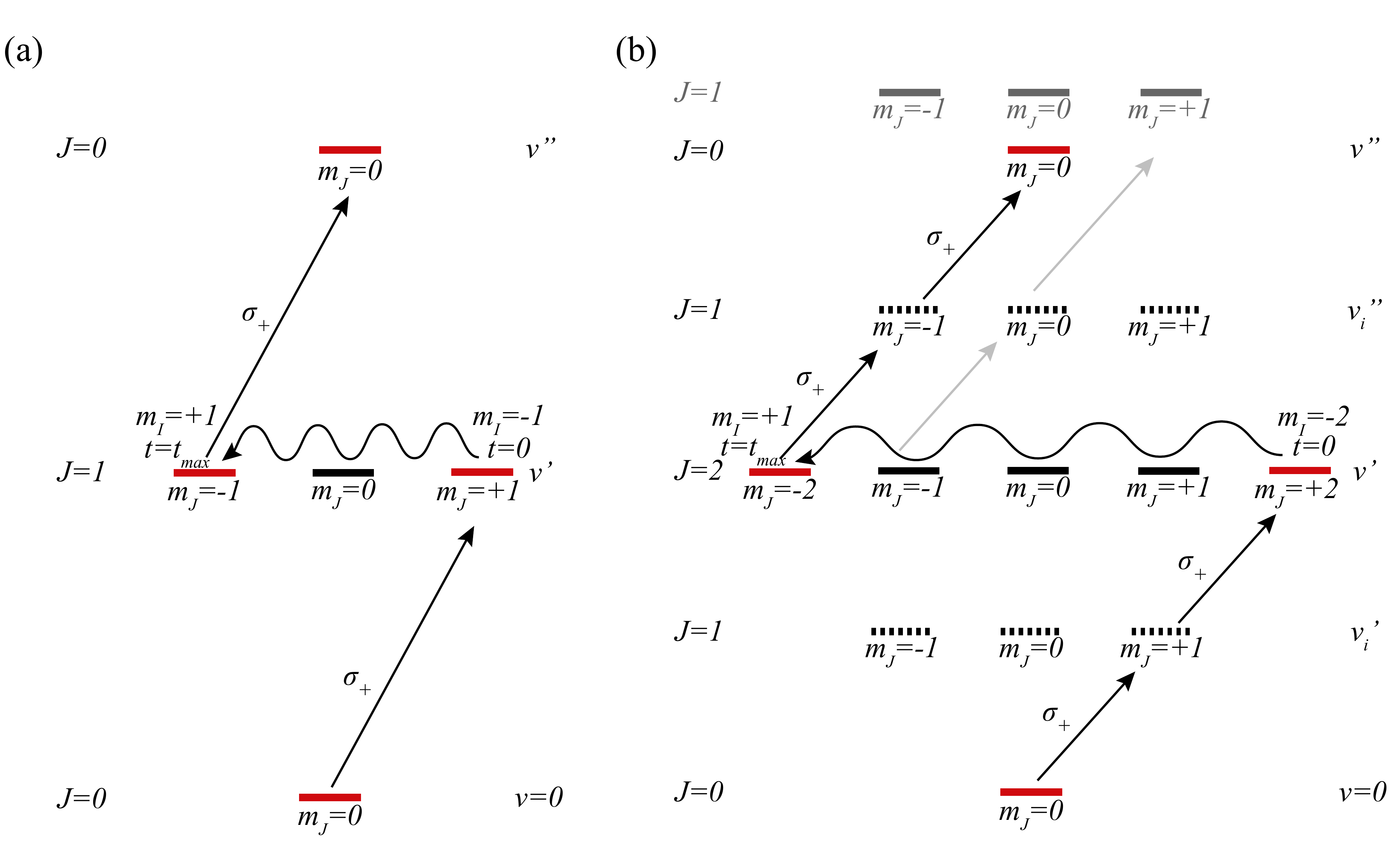}
	\caption{Excitation schemes for production of nuclear spin-polarized molecules with (a) $I=1$ (e.g. with constituent atoms of D or H\textsubscript{2}), and (b) $I=3/2$ or $2$ (e.g. for HD, DT or D\textsubscript{2}). The squiggly arrows show rotational polarization reduction $\Delta m_J = -2$ or $-4$, associated with nuclear spin increase $\Delta m_I = +2$ or $+4$ (for Figs.~\ref{fig:sketches}(a) or~\ref{fig:sketches}(b), respectively). Only 100\% nuclear-spin-polarized molecules reach the upper state. \label{fig:sketches}}	
\end{figure}

We describe in more detail why this method should surpass conventional methods by several orders of magnitude. Conventional molecular-beam separation methods are based on the differential deflection of the spin-projection states, from a beam with a cross sectional area $A$, velocity $v$, and density $\rho$, with large, localized B-field or E-field gradients, which limit ${A}{\sim}1$ cm\textsuperscript{2} to be small and localized. However, the beam density must be low enough so that the beam divergence, from collisions, is less than the differential deflections of the target spin-projection states. This constraint limits the beam density to ${\rho}{\sim}10^{11}$ cm\textsuperscript{-3} for open-shell atoms~\cite{nass}. The beam separation occurs over a distance of ${\sim}1$ m and on the ms timescale, with a beam velocity $v$ typically in the range $0.2{-}2.5 {\times} 10^5$ cm/s. The largest production rate (given by $Av\rho$) for spin-polarized H atoms, which is the most favorable case, has been ${\sim}5{\times}10^{16}$ H/s~\cite{nass}. For closed-shell molecules, the production rates are several orders of magnitude lower (as nuclear magnetic moments are ${\sim}10^3$ times smaller than electronic ones, due to the ${\sim}10^3$ larger mass).  

\begin{figure}
	\includegraphics*[width=1\textwidth]{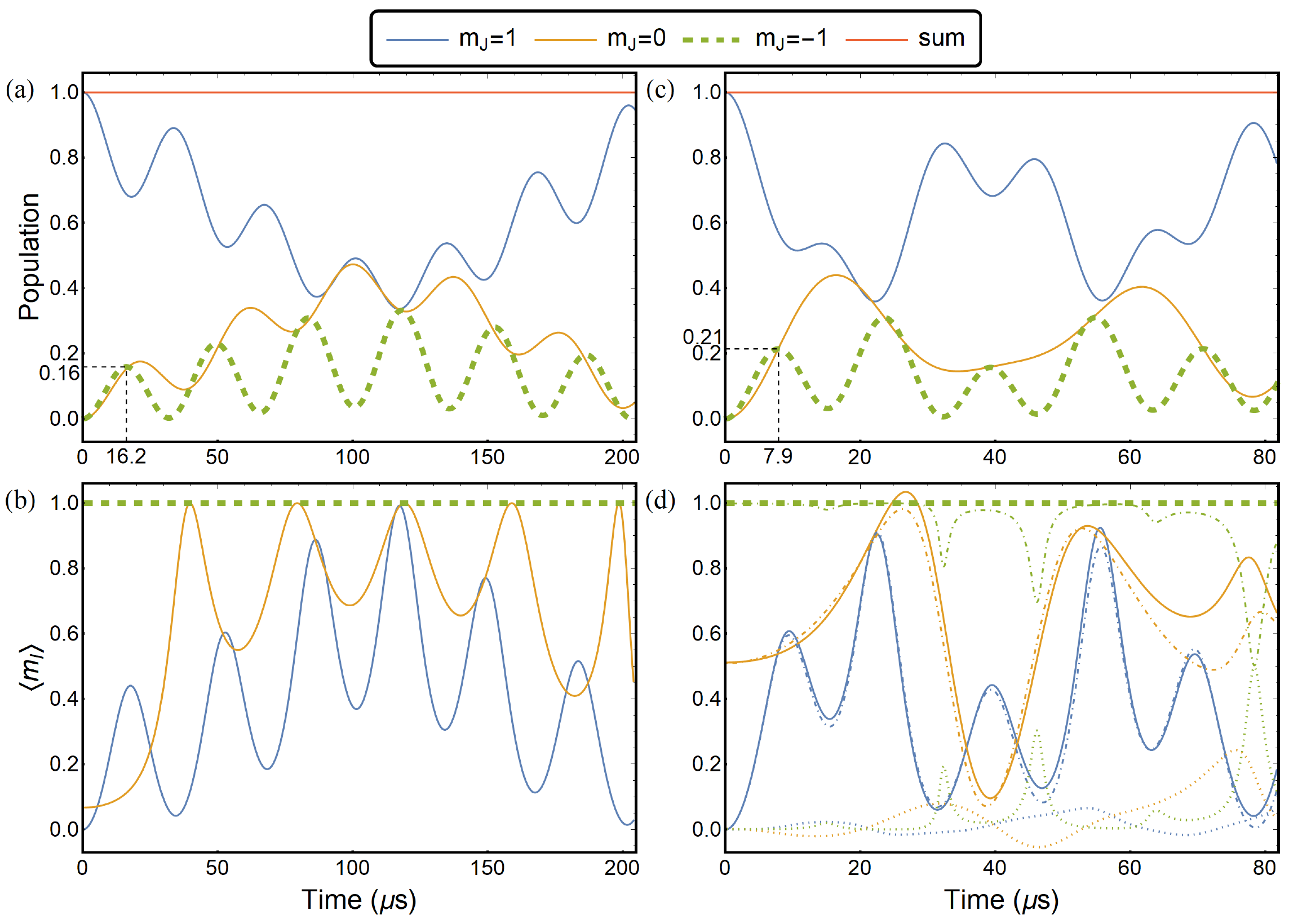}
	\caption{CH\textsubscript{2}O (a,b): Population and $\langle m_I \rangle$ of $m_J$ states as function of time. The $m_J = -1$ state  reaches 15.9\% of the total population at ${\sim}16.2$ $\rm{\mu s}$. CHDO (c,d): Population and $\langle m_I \rangle$ of $m_J$ states as function of time. The solid lines (and the dashed thick line) correspond to the total nuclear spin projection, whereas the dash-dotted and dotted lines correspond to $\langle m_{I_D} \rangle$, and $\langle m_{I_H} \rangle$ respectively. The $m_J = -1$ state  reaches 21.4\% of the total population at 7.9 $\rm{\mu s}$. By that time, mainly the deuteron is polarized. \label{fig:CH2O}}
\end{figure}

In contrast, the polarization of nuclear spins in a molecular beam, via the IR-excitation method, occurs by the intramolecular transfer of rotational to nuclear polarization, via the hyperfine interaction on the ${\rm \mu s}$ timescale, and thus does not require any type of beam separation or external inhomogeneous static fields. The lack of need to spatially separate the spin-states in the molecular beam, the shorter polarization time,  and the ability to use a long slit expansion (not possible for conventional beam-separation methods due to the inhomogeneous fields), allow the beam area $A$ and density $\rho$ to be much higher than for beam-separation methods, e.g. $A{\sim}$10-100 cm\textsuperscript{2}, and $\rho{\sim}10^{15}$ cm\textsuperscript{-3}, which are 1-2 and 4 orders of magnitude higher, respectively. This increase in $A$ and $\rho$ potentially allows a total production rate $Av\rho$ of up to ${\sim}10^{22}$ s\textsuperscript{-1}. The production rate is limited by the absorbed photons for the IR-excitation and photodissociation steps: tabletop IR-excitation and photodissociation lasers producing ${\sim}10^{21}$ photons s\textsuperscript{-1} are available commercially; industrial-scale lasers can likely reach higher. We note that related arguments were used to predict the production of ultrahigh densities of spin-polarized hydrogen (SPH) atoms from hydrogen halide photodissociation~\cite{sofikitis17}, and indeed $10^{19}$-$10^{20}$ SPH were observed~\cite{sofikitis18,spiliotis2021,spiliotis2021depolarization}, 8-9 orders higher than those from beam-separation methods; therefore surpassing conventional beam-separation methods by many orders of magnitude is not unprecedented.

\section{Production of spin-polarized H\textsubscript{2} (isotopes) from formaldehyde IR excitation and photodissociation}\label{sec2}

The IR-excitation method depends crucially on transferring a large fraction of the population to the 100\% nuclear-spin-polarized state, and in short enough time ($t_{max} \lesssim 30 \, {\rm \mu s}$) to avoid significant depolarization under these conditions. Below, we give a description of the calculation of the polarization transfer and $t_{max}$ for CH\textsubscript{2}O and CH\textsubscript{2}O\textsubscript{2} isotopes.

The CH\textsubscript{2}O ortho states ($I=1$) are described by $K_{-1} = 1,3,5,\dots$ and the para states by $K_{-1} = 0,2,4,\dots$ in the notation $J_{K_{-1}K_{1}}$~\cite{cross}. The hyperfine hamiltonian matrix elements of the $J=1,\, K_{-1} =1$ state can be written as~\cite{ch2o}:
\begin{equation}\label{eqch2o}
\begin{split}
E/h =& C (-1)^{I + J +F} [I(I+1)(2I+1)J(J+1)(2J+1)]^{1/2}\\
&\times\left\{\begin{array}{ccc}
F & I & J \\
1 & J & I  
\end{array}\right\}
+ D (-1)^{I + J +F} \left\{\begin{array}{ccc}
F & I & J \\
2 & J & I   
\end{array}\right\}\\ &\times\Bigg[\frac{(I+1)(2I+1)(2I+3)J(J+1)(2J+1)}{I(2I-1)(2J-1)(2J+3)}\Bigg]^{1/2} , 
\end{split}
\end{equation}
where $I=1$ is the sum of proton spins, $J$ the rotational angular momentum, $F=J+I$ the total angular momentum, $C$ the spin-rotation interaction constant and $D$ the spin-spin interaction constant. Substituting $C(1_{11})=-3.07$ kHz and $D(1_{11})=8.87$ kHz~\cite{ch2o} for the lower state $K_{1} =1$ we obtain the hamiltonian matrix elements in the $|F,\, m_F \rangle$ representation. After transforming to the $|m_J , \, m_I \rangle$ basis as:
\begin{equation}\label{eqbt}
|m_J , \, m_I \rangle = \sum_{F,\, m_F} \langle F, \, m_F | m_J , \, m_I \rangle |F, \, m_F \rangle,
\end{equation}
the hamiltonian is diagonalized to obtain the time evolved states~\cite{kannis} shown in Fig.~\ref{fig:CH2O}, where at $t_{max} = 16.2$ ${\rm \mu s}$, 15.9\% of the population has been transferred to the $m_J = -1$ state, which is always fully nuclear-spin-polarized (Fig.~\ref{fig:CH2O}(b)). The $\langle m_I \rangle$ of each $m_J$ state is plotted (Fig.~\ref{fig:CH2O}(c,d)), giving the degree of nuclear polarization for each state.

\begin{figure}
	\includegraphics*[width=1\textwidth]{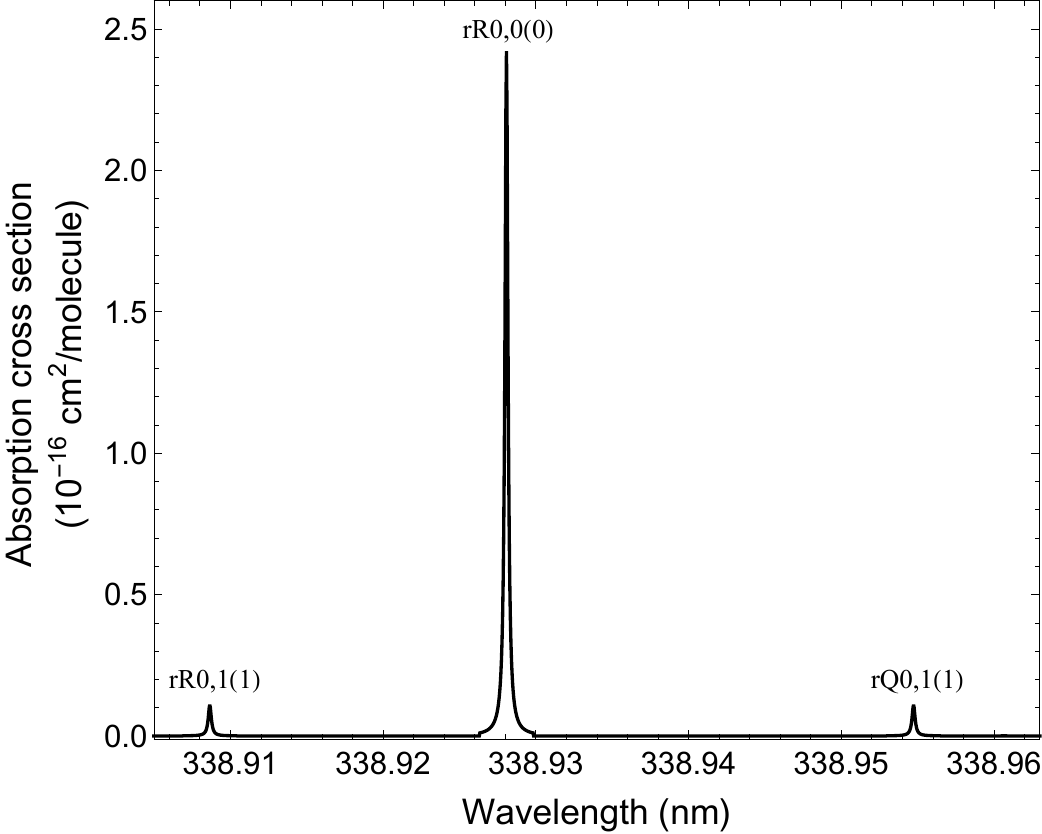}
	\caption{Simulated absorption spectrum of the $2^1 _0 4^1 _0$ vibronic band of formaldehyde, using Gaussian and Lorentzian FWHM of 0.003 cm\textsuperscript{-1} and 0.02 cm\textsuperscript{-1}, respectively, and $T=1$ K. Relevant transitions are labeled. The absorption spectrum simulation of the $2^1 _0 4^1 _0$ vibronic band was performed using the PGOPHER~\cite{pgopher} program and the supporting material provided by~\cite{orrewing}. \label{fig:CH2Odiss}}
\end{figure}

The same technique can be applied to singly deuterated formaldehyde, CHDO. The effective hyperfine hamiltonian for asymmetric top molecules with two (nonzero) nuclear spins is written as~\cite{thaddeus}:
\begin{equation}\label{eqchdo}
\begin{split}
H/h = &\frac{(eq_J Q)_D}{2 I_D (2 I_D -1) J (2 J - 1)} \Big[ 3 (\vec{I}_D \cdot \vec{J})^2 +\frac{3}{2}(\vec{I}_D \cdot \vec{J})\\
&-\vec{I}_D ^2 \vec{J} ^2 \Big] +\frac{d'_J}{J (2 J-1)}\Big[ \frac{3}{2} (\vec{I}_D \cdot \vec{J})(\vec{I}_H \cdot \vec{J})\\ &+\frac{3}{2}(\vec{I}_H \cdot \vec{J})(\vec{I}_D \cdot \vec{J})
-(\vec{I}_H \cdot \vec{I}_D)\vec{J}^2 \Big] \\
&+ C_H \vec{I}_H \cdot \vec{J} + C_D \vec{I}_D \cdot \vec{J},
\end{split}
\end{equation}
where $(eq_J Q)_D,\, d'_J,\, C_H, \, {\rm{and}}\, C_D$ are the quadrupole interaction coupling constant, the nuclear spin-spin interaction coupling constant, the proton and deuteron spin-rotation interaction coupling constants, defined in~\cite{thaddeus}.

The hyperfine structure of the $1_{11}$ and $1_{10}$ states of CHDO has been studied by Tucker et al.~\cite{tucker}. We investigate the polarization dynamics of the $1_{10}$. Substituting $C_H = -1.43$ kHz, $C_D = 0.11$ kHz given in~\cite{tucker}, $d'_J = -0.54$ kHz which is calculated from molecular geometry (as it is presented in Fig. 2 of~\cite{flygare}), and $(eq_J Q)_D = 17$ kHz, where the $\chi_{aa}$, $\chi_{bb}$, and $\chi_{cc}$ are taken from Table V of~\cite{flygare}. Following the polarization scheme presented in Fig.~\ref{fig:sketches}(a), we achieve 21.4\% of population with 100\% deuteron polarization at 7.9 $\rm{\mu s}$ (see Fig.~\ref{fig:CH2O}(c,d)).

To produce nuclear-spin-polarized HD or DT molecules, a different scheme (Fig.~\ref{fig:sketches}) is required. Starting with $m_J {=}2$ ($t{=}0$), at time $t{=}t_{max}$ the population of the $m_J {=} -1$ state is maximized (which has 100\% nuclear-spin-polarization), and is then excited to the $J{=}1$, $m_J {=} +1$ state. For the calculation of $t_{max}$ the hyperfine constants for the $J{=}2$ state are needed. However, the $C_{H,D}$ values found in the literature~\cite{thaddeus,flygare,tucker} are not consistent and the $M_{gg} ^{H,D}$ have not been measured. Thus, we cannot accurately determine $t_{max}$ unless all the hyperfine constants of CHDO are known. The proton spin-rotation constants are particularly required, because they are mainly responsible for the polarization transfer to the proton. In the case of D, the corresponding quadrupole coupling interaction is dominating.

\section{ Production of spin-polarized H\textsubscript{2} (isotopes) from formic acid IR excitation and photodissociation}
\begin{figure}
	\centering
	\includegraphics*[width=1\textwidth]{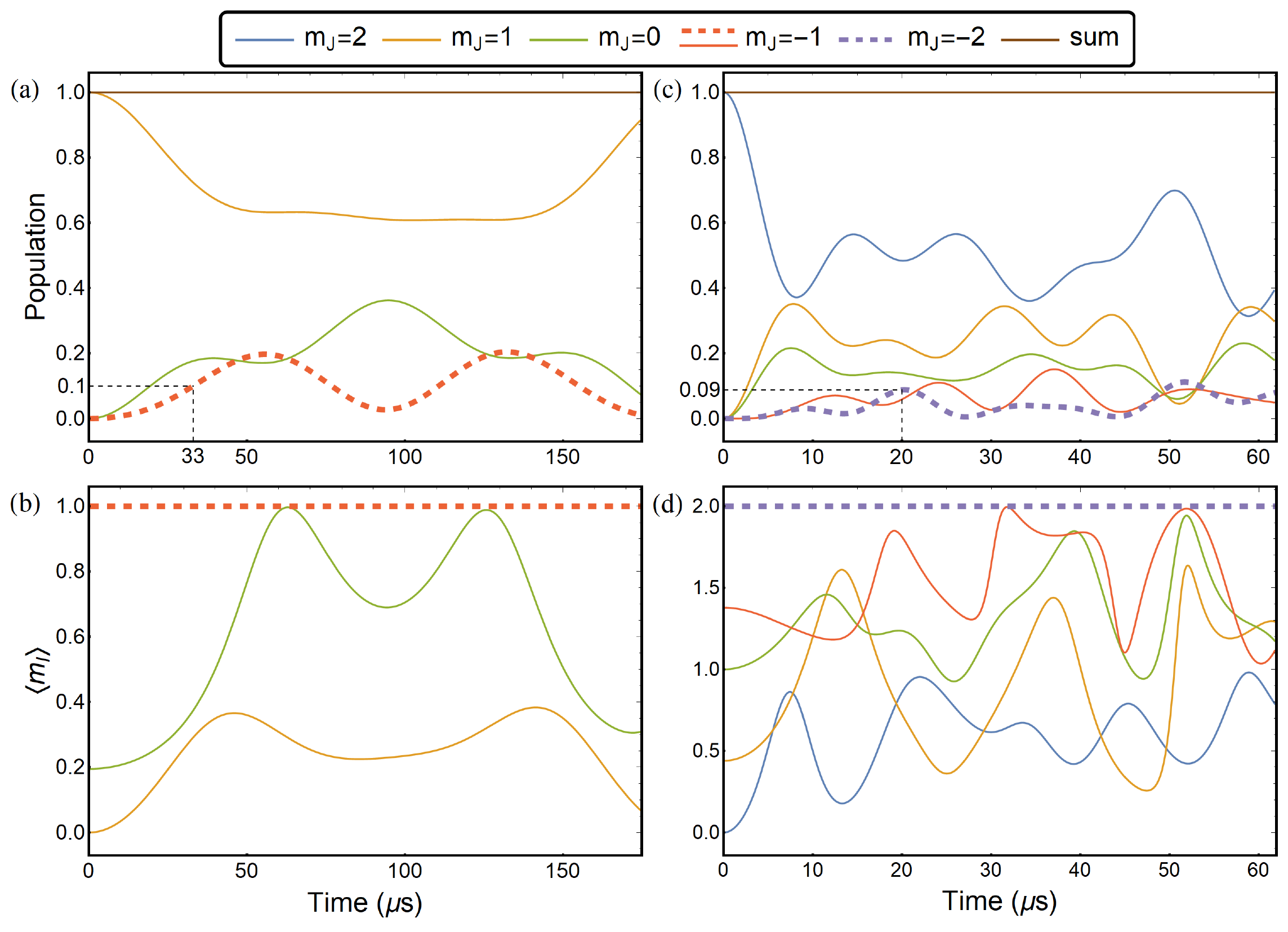}
	\caption{CH\textsubscript{2}O\textsubscript{2}(a,b): Population and total $\langle m_I \rangle$ of $m_J$ states as function of time. The $m_J = -1$ state  reaches 10\% of the total population at $33$ ${\rm\mu s}$. CD\textsubscript{2}O\textsubscript{2}(c,d): Population and total $\langle m_I \rangle$ of $m_J$ states as function of time. The $m_J = -2$ state  reaches 8.8\% of the total population at ${\sim}20$ ${\rm\mu s}$. \label{fig:HCOOH}}
\end{figure}

Similar methods can be applied to ortho-trans-formic acid and isotopes. The effective hyperfine hamiltonian of formic acid (CH\textsubscript{2}O\textsubscript{2}) is written as~\cite{thaddeus,chardon,cazzoli}:
\begin{equation}\label{eq:formham}
\begin{split}
H/h =& C_{H_1} \vec{I}_1 \cdot \vec{J} + C_{H_2} \vec{I}_2 \cdot \vec{J}\\
&+ \frac{d'_J}{J (2J-1)}\Big[\frac{3}{2} (\vec{I}_1 \cdot \vec{J}) (\vec{I}_2 \cdot \vec{J})\\
&+ \frac{3}{2}(\vec{I}_2\cdot\vec{J})(\vec{I}_1\cdot\vec{J})
-(\vec{I}_1 \cdot \vec{I}_2 )\vec{J}^2 \Big],
\end{split}
\end{equation}
where $I_1 = I_2 = 1/2$, $d'_J = -\frac{6\mu_N ^2 g_H ^2}{(J+1)(2J +3) r^3} \sum_{g} \frac{r_{g} ^2}{r^2} \langle J_g ^2\rangle -\frac{1}{3} J (J+1)$, $C_{H_{1,2}} = \sum_{g} \frac{M_{gg} ^{1,2} \langle J_g ^2\rangle}{J (J+1)}$, and the constants labeled with 1 refer to the proton bounded to the carbon atom. For the $1_{11}$ state, $C_{H_1} = -3.8$ kHz, $C_{H_2} = -4.1$ kHz~\cite{cazzoli}, and $d'_J = 1.8$ kHz~\cite{bellet,hocking}. Figure~\ref{fig:HCOOH} shows that the population of the $m_J =-1$ state reaches 10\% at $t{\sim}33$ ${\rm\mu s}$, assuming 100\% rotational polarization at $t=0$.

The same technique, but for the $|\nu^{\prime} ,\, J=2\rangle$ state (Fig.~\ref{fig:sketches}(b)) can be applied to double deuterated formic acid (CD\textsubscript{2}O\textsubscript{2}). The total wavefunction is symmetric in the exchange of nuclear spins $I_1$ and $I_2$. In the case of $K_{-1}={\rm even}$ states, the rotation wavefunction is symmetric, so the nuclear spin wavefunction must be symmetric, i.e. $I=I_1 +I_2 = 0,2$ (ortho states). Therefore, the non-zero hamiltonian matrix elements in the $| J, \, I, \, F \rangle$ representation are~\cite{flygare}:
	\begin{equation}\label{eqd2}
	\begin{split}
	\langle J,\, I, \, F| H/h | J,\, I,\, F \rangle =&C_D \frac{A}{2 J (J + 1 )}+\Big[ a_Q \frac{3 I (I+1) - 11}{(2I-1)(2I+3)}\\
	&+ a_S \frac{I(I+1) +8}{2 (2I-1)(2I+3)} \Big] \frac{3A(A-1)-4I(I+1) J (J+1)}{2J(J+1)(2J-1)(2J+3)},\\
	\langle J,\, I, \, F| H/h | J,\, I-2,\, F \rangle =& \frac{3(2 a_Q - a_S)}{8J(J+1)(2J-1)(2J+3)(2I-1)}\\
	&\times \sqrt{\frac{(I+3)(3-I)(I+2)(4-I)}{(2I-3)(2I+1)}} [(J-F+I)(J+F+I+1)\\
	&\times (F+I-J)(J+F-I+1)(J-F+I-1)(J+F+I)\\
	&\times(F+I-1-J)(J+F-I+2)]^{1/2} ,
	\end{split}
	\end{equation}
where $A= J(J+1) + I(I+1) - F(F+1)$. The constants $a_Q,\, a_S,\,{\rm{and}}\, C_D$ are the quadrupole interaction coupling constant, the nuclear spin-spin interaction coupling constant, and the spin-rotation interaction coupling constant, defined in~\cite{flygare}.

Cazzoli et al.~\cite{cazzolidcood} measured the hyperfine constants $\chi_{gg}$ and calculated $M_{gg}$ (see cc-pCVQZ calculation) using a semi-experimental equilibrium geometry. Figure~\ref{fig:HCOOH}(c,d) shows the hyperfine beats of the $2_{21}$ state which is initially rotationally polarized. The values of the hyperfine constants are $a_Q=-152.85$ kHz, $a_S = 0.24$ kHz, and $C_D = -2.85$ kHz. The 8.8\% of the population has been transferred to the $m_J = -2$ state, at $t_{max} = 20$ ${\rm \mu s}$. This state is always fully nuclear-spin-polarized (Fig.~\ref{fig:HCOOH}(c,d)).

We have described the production of spin-polarized molecular photofragments from the IR-excitation and photodissociation of molecular beams, with rates that may approach the IR-photon production rates of ${\rm 10^{21}\, photons\, s^{-1}}$. Details on the production of spin-polarized H\textsubscript{2} and H\textsubscript{2}O isotopes, from CH\textsubscript{2}O and CH\textsubscript{2}O\textsubscript{2}, have been given. The production rates of such method are mainly limited by the efficiency of the IR-excitation, and UV-photodissociation, and trapping steps. We have given arguments that the efficiencies of these steps can be high, however since the scale is so far beyond what has already been demonstrated, proof-of-principle demonstrations of these are necessary to verify to what extent they can be achieved. The point of this paper is to motivate such experimental demonstrations. In addition, future improvements in the power of IR and UV lasers will allow further increases in the production rates of this method. 

This method of macroscopic production of polarized molecules can be generalized to several other systems, such as for spin-polarized O\textsubscript{2}, N\textsubscript{2}, NO, and \textsuperscript{13}CO, from the IR-excitation and photodissociation of O\textsubscript{3}, N\textsubscript{2}O, NO\textsubscript{2},  and CO\textsubscript{2}.

This work is supported in part by the Hellenic Foundation for Research and Innovation (HFRI) and the General Secretariat for Research and Technology (GSRT), through the grant agreement No. HFRI-FM17-3709 (project NUPOL).

%\bibliography{MNHFP_biblio}

\end{document}